\documentclass[twocolumn,secnumarabic,amssymb, nobibnotes, aps, prd]{revtex4}
\usepackage{graphicx}

\begin{document}
\title{What is the azimuthal quantum force in superconductor}
\author{A. V. Nikulov}
\affiliation{Institute of Microelectronics Technology and High Purity Materials, Russian Academy of Sciences, 142432 Chernogolovka, Moscow District, RUSSIA.} 
\begin{abstract}J.E. Hirsch notes in the recent paper arXiv: 0908.409 that the azimuthal quantum force, introduced ten years ago, would violate the principle of angular momentum conservation. It is specified in the present work that the essence and the method of logical deduction of the quantum force do not overstep the limits of the universally recognised quantum formalism. The puzzle revealed by J.E. Hirsch concerns rather of quantum mechanics in whole than a certain theory or work. The essence of this and other puzzles generated with quantum mechanics is considered shortly. 
 \end{abstract}

\maketitle

\narrowtext

\section*{Introduction}
\label{intro}
The Meissner effect, interpreted fairly as the first experimental evidence of macroscopic quantum phenomenon, was discovered as far back as 1933. But, as J.E. Hirsch notes in the recent paper \cite{Hirsch2010}, {\it "Strangely, the question of what is the 'force' propelling the mobile charge carriers and the ions in the superconductor to move in direction opposite to the electromagnetic force in the Meissner effect was essentially never raised nor answered to my knowledge, except for the following instances: \cite{London1935}  (H. London states: "The generation of current in the part which becomes supraconductive takes place without any assistance of an electric field and is only due to forces which come from the decrease of the free energy caused by the phase transformation," but does not discuss the nature of these forces), \cite{PRB2001} (A.V. Nikulov introduces a "quantum force" to explain the Little-Parks effect in superconductors and proposes that it also explains the Meissner effect)"}. I would like in this paper to define more exactly the essence of the azimuthal quantum force introduced ten years ago \cite{PRB2001}. This quantum force can not explain the Meissner effect. Moreover the utilization of the quantum force in \cite{PRB2001} does not overstep the limits of the universally recognised quantum formalism and it can rather describe than explain quantum phenomena as well as quantum mechanics in whole. 

\section{The description of the quantization effects in superconductors }
\label{sec:1}
The Meissner effect is considered as a particular case of quantization in the universally recognised description. According to the Ginzburg-Landau theory \cite{GL1950} superconducting state can be described with help of the GL wave function $\Psi _{GL} = |\Psi _{GL}|\exp i\varphi $, in which $|\Psi _{GL}|^{2} = n_{s}$ is interpreted as the density of superconducting pairs and $\hbar \bigtriangledown \varphi = p = mv + qA$ is momentum of single pair with the mass $m$ and the charge $q = 2e$. Superconducting current density is 
$$j_{s} = \frac{n_{s}q}{m}(\bigtriangledown \varphi - qA) = n_{s}q v \eqno{(1)}$$ 
according to the GL theory, when the pair density is constant in the superconductor $\bigtriangledown n_{s} = 0$ \cite{Tinkham}. The quantization can be deduced from the requirement that the complex wave function must be single-valued $\Psi _{GL} = |\Psi _{GL}|\exp i\varphi = |\Psi _{GL}|\exp i(\varphi + n2\pi)$ at any point in superconductor \cite{Tinkham}. Therefore, its phase must change by integral multiples of $2\pi $ following a complete turn along the path of integration, yielding the Bohr-Sommerfeld quantization 
$$\oint_{l}dl \nabla \varphi = \oint_{l}dl \frac{p}{\hbar}  = \oint_{l}dl \frac{mv + qA}{\hbar} = n2\pi \eqno{(2)}$$ 
According to the relations (1), (2) and  $\oint_{l}dl A = \Phi $ the integral of the current density along any closed path inside superconductor 
$$\mu _{0}\oint_{l}dl \lambda _{L}^{2} j_{s}  + \Phi = n\Phi_{0}  \eqno{(3)}$$  
must be connected with the integral quantum number $n$ and the magnetic flux $\Phi $ inside the closed path $l$. $\lambda _{L} = (m/\mu _{0}q^{2}n_{s})^{0.5} = \lambda _{L}(0)(1 - T/T_{c})^{-1/2}$  is the London penetration depth; $\lambda _{L}(0) \approx 50 \ nm = 5 \ 10^{-8} \ m$ for most superconductors \cite{Tinkham}; $\Phi _{0} = 2\pi \hbar /q$ is the flux quantum.

The relation (3) can describe the Meissner effect \cite{Meissner1933}, the magnetic flux quantization observed first in 1961 \cite{FQ1961} and the quantum periodicity in different parameters of superconducting cylinder or ring with narrow wall $w \ll \lambda _{L}$. According to the quantization relation (3), the persistent current \cite{PC1961}
$$I_{p} = sj_{s} = \frac{q\hbar}{mr\overline{(s n _{s})^{-1}}} (n - \frac{\Phi }{\Phi _{0}}) = I_{p,A}2 (n - \frac{\Phi }{\Phi _{0}})  \eqno{(4)}$$  
must flow along the narrow wall $w \ll \lambda _{L}$ of cylinder or ring with a radius $r$, section $s = wh > 0$ and pair density $n _{s} > 0$  along the whole circumference $l$, when magnetic flux inside the cylinder or ring $\Phi = BS = B\pi r^{2}$ is not divisible the flux quantum $\Phi \neq n\Phi _{0}$. Here $h$ and $w$ are the height and width of ring or cylinder wall; $\overline{(s n _{s})^{-1}} = l^{-1}\oint _{l}dl (s n _{s})^{-1}$ is the value determining the amplitude $I _{p,A} = q \hbar/2mr\overline{(s n _{s})^{-1}} $  of the persistent current in a ring with section $s$ and density $ n _{s}$ which may vary along the ring circumference $l$ \cite{PRB2001}.  The quantum periodicity is observed because of the change with magnetic flux value $\Phi $ of the integer quantum number $n$ corresponding to the minimal energy $\propto (n - \Phi /\Phi _{0})^{2} $. This phenomenon was observed first as far back as 1962 by W. A. Little and R. D. Parks \cite{LP1962} at measurements of the resistance of thin cylinder in the temperature region corresponding to its superconducting resistive transition. Later on, the quantum oscillations of the ring resistance $\Delta R \propto  I_{p}^{2}$ \cite{Letter07,toKulik2010}, its magnetic susceptibility $\Delta \Phi _{Ip} = LI_{p}$ \cite{PCScien07}, the critical current $I_{c}(\Phi /\Phi_{0}) = I_{c0} - 2|I_{p}(\Phi /\Phi_{0})| $ \cite{JETP07J} and the dc voltage $V_{dc}(\Phi /\Phi_{0}) \propto I_{p}(\Phi /\Phi_{0})$ measured on segments of asymmetric rings \cite{Letter07,toKulik2010,PerMob2001,Letter2003,PCJETP07} were observed. 

The magnetic flux quantization is observed \cite{FQ1961} in superconducting cylinder or ring with wide wall $w \gg \lambda _{L}$. The magnetic flux should have the discrete values $\Phi = n\Phi_{0} $ according to (3) because of zero current density $j_{s} = 0$ along a circuit $l$ inside the wide wall. The path of integration $l$ in (2) and (3) can be tightened in point without the quantum number $n$ change if the wave function is valid in whole region inside it. Therefore the Meissner effect $\Phi = n\Phi_{0} = 0$ can be describe as a consequence of the requirement that the quantum number $n \neq 0$ in (2) and (3) only if the wave function $\Psi _{GL} = |\Psi _{GL}|\exp i\varphi $ has a singularity inside $l$, i.e. there is a non-superconducting region. The Abrikosov vortices, singularities with $n = 1$, $\oint_{l}dl \nabla \varphi = 2 \pi $, make it possible for magnetic flux to penetrate inside superconductor without total destruction of superconductivity. Both the Meissner effect and the Abrikosov state are marvellous experimental evidence of long-rang order of phase coherence \cite{Vort2003}. Numerous observations of the Abrikosov state \cite{Huebener} corroborate the relation $\Phi = n\Phi_{0} $ between the magnetic flux $\Phi $ and the number $n$ of the Abrikosov vortices inside a macroscopic circuit $l$ and thus, the long-rang phase coherence (2). 
 
\section{Transition between discrete and continuous spectrum of superconducting ring states}
\label{sec:2}
There is important to accentuate that the quantum number $n$ is connected definitely (3) with measurable variables, for example, the persistent current $I _{p}$ and the magnetic flux $\Phi = BS = B\pi r^{2}$ inside a narrow wall ring (4), when the path of integration $l$ in (2) can be closed, i.e. when the phase coherence is valid along the whole ring circuit. According to the quantization condition (3) the persistent current $I _{p} = s2en _{s}v$ (4) and the pair velocity $v  = I _{p}/sqn _{s}$ should increase with the $\Phi $ value until the change of the quantum number $n = \oint _{l}dl \nabla \varphi /2\pi $. The $n$ change means the jump of the persistent current on the value $ 2I _{p,A}$ (4), of the pair velocity on $v _{n} - v _{n+1} = 2I _{p,A}/sqn _{s}$ and the angular momentum $pr = \oint _{l}dl p/2\pi $ of each pair on the Planck constant $\hbar $ (2). These jumps, Fig.1, observed at the measurements of the ring magnetization $\Delta \Phi = LI _{p}$ below superconducting transition $T < T _{c}$ \cite{nChGeim,nChMoler} can be described as a consequence of the transition in the normal state with $n _{s} = 0$,  $R _{A} > 0$ at least of one ring segment $l _{A}$, Fig.2. The electric current circulating in the ring with non-zero resistance $ R _{A}  > 0$ should decrease $I(t) = I_{p}\exp -t/\tau_{RL}$ from $I _{p} \approx  2 I _{p,A} \times  1.8$ to $I _{p} \approx  2 I _{p,A} \times  0.8$ at the quantum number change from $n = 0$ to $n = 1$, Fig.1, during very short time $t \approx  \tau_{RL} \ln (1.8/0.8)$ equal in order of value the relaxation time $\tau_{RL} = L/R _{A}$. The angular momentum of mobile charge carriers $M _{p} = (2m/q)I _{p}S = (2m/q)I _{p,A}\pi r^{2}2(n - \Phi /\Phi _{0}) = \hbar (2\pi r/ \overline{(s n _{s})^{-1}})$ changes from $M _{p} \approx  \hbar N _{s} \times 1.8$ to $M _{p} \approx  \hbar N _{s} \times 0.8$ during this time. Here $2\pi r/ \overline{(s n _{s})^{-1}} \approx 2\pi r s n _{s} = N _{s}$ is the number of superconducting pairs in the homogeneous state of the ring, when  $s$ and $ n _{s}$ does not change along $l$. The change on the macroscopic value $\Delta M _{p} \approx  \hbar N _{s}$ takes place under the influence of the real forces: the dissipation force $F _{dis} = -\eta \overline{v} $ acting on electrons in the normal $l _{A}$ segment induces a potential difference $ V_{A} = R_{A}I(t)= R_{A}I_{p}\exp {-t /\tau_{RL}}$ the electric field $E = - \bigtriangledown V_{A}$ of which maintains $E \approx -V_{A}/ l_{A}$ the current $I(t) $ of normal electrons in this segment at the force balance $F _{dis} + qE = 0$ and brakes $E \approx V_{A}/(l-l_{A})$ superconducting pairs in the superconducting segment $l-l_{A}$ in accordance with the Newton's second law $mdv/dt = qE$.

\begin{figure}[]
\includegraphics{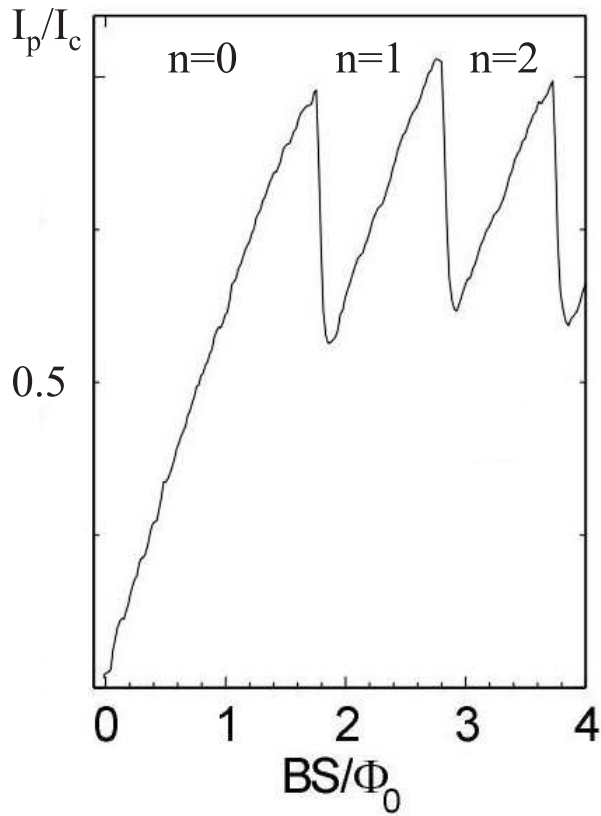}
\caption{\label{fig:epsart} Dependence of the persistent current $I_{p}$ of aluminium ring with the radius $r \approx  0.7 \ \mu m$ and section $s \approx  0.01 \ \mu m^{2}$ on the external magnetic field $B$ obtained with help of the magnetization $\Delta \Phi = LI _{p}$ measurement at the temperature $T \approx  0.6 \ K \approx  0.5T _{c}$. The dependence $I_{p}(B)  = I_{p,A}2 (n - (BS+LI_{p})/\Phi _{0})$ is not quite linear because of the additional magnetic flux $\Delta \Phi = LI _{p}$ equal approximately $LI _{c} \approx  0.3\Phi _{0}$ at the maximal value of the persistent current $I _{p} \approx  I _{c} \approx  300 \ \mu A$. This $I _{p} $ value corresponds to the macroscopic magnetic moment $M _{m} = I _{p}S \approx  5 \ 10^{8} \ \mu _{B}$ and angular momentum $M _{p} = (2m/e) M _{m} \approx  5 \ 10^{8} \ \hbar \approx N _{s} \times 1.8 \hbar $ of $N _{s} \approx  3 \ 10^{8} $ pairs in the ring, where $\mu _{B}$ is the Bohr's magneton and $\hbar$ is the reduced Planck constant.}
\end{figure}

\begin{figure}[b]
\includegraphics{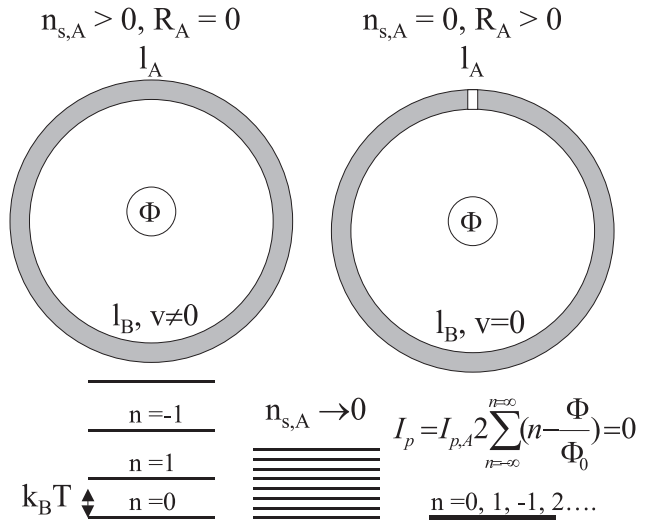}
\caption{\label{fig:epsart}The strongly discrete spectrum $ \Delta E_{n,n+1} \gg k _{B}T$ of permitted states of a superconducting ring (the left ring) becomes continuous at the transition of any ring segment $l _{A}$ in the normal state (the right ring). For example, at $\Phi = \Phi _{0}/4$ the energy difference $ E_{n= 1} - E_{n= 0} \gg k _{B}T$,  $ E_{n= -1} - E_{n= 1} \gg k _{B}T$ $\cdot \cdot \cdot $ when $n _{s,A} > 0$ (the left ring) but $ E_{n+ 1} - E_{n} = 0 \ll k _{B}T$ when $n _{s,A} = 0$ (the right ring). The reality of this qualitative change, possible thanks to the reality of the density $|\Psi _{GL}|^{2} = n_{s}$ described with the GL wave function $\Psi _{GL} = |\Psi _{GL}|\exp i\varphi $, generates some puzzles. For example, the questions: "Why and how quickly can the velocity of pairs in a segment $l _{B}$ change from $v = 0$ to $v \neq 0$ after the transition of the $l _{A}$ segment into the superconducting state $n _{s,A} > 0$.}
\end{figure}

There is important to note that the decrease of the pair density $n _{s,A}$ in a ring segment down to zero is qualitative transition from discrete to continuous spectrum of superconducting ring states, Fig.2. The persistent current $I _{p} = sn _{s}qv _{n}$ (4) increases the energy of superconducting state of a ring on the kinetic energy 
$$E_{n} = \oint _{l} dl sn _{s} \frac{mv _{n}^{2}}{2} =  \frac{I_{p}}{q}\oint _{l} dl \frac{mv_{n}}{2} = I _{p,A}\Phi _{0} (n - \frac{\Phi }{\Phi _{0}})^{2}  \eqno{(5)}$$
of pairs. This increase (5) is considered as the cause of the Little - Parks effect \cite{Tinkham,LP1962}. The spectrum of the permitted states (5) is strongly discrete $ \Delta E_{n,n+1} = |E_{n+1} - E_{n}| \gg k _{B}T$ thanks to the enormous number $ N _{s} = \oint _{l} dl sn _{s}$ of pairs in a real superconducting ring, see, for example, the caption of Fig.1. The energy difference between permitted states $ \Delta E_{n,n+1} \approx  I _{p,A} \Phi _{0} \approx  4 \ 10^{-21} \ J$ (5) at the real amplitude $ I _{p,A} \approx  2 \ \mu A$ of the persistent current of a real ring measured even near superconducting transition $T \approx  0.99T _{c} \approx  1.24 \ K$ \cite{JETP07J} corresponds to the value $ \Delta E_{n,n+1} / k _{B} \approx  300 \ K $ exceeding strongly the temperature of measurements $T \approx 1.24 \ K$.  The observations only single state $n$ at $(n - 0.5)\Phi _{0} < \Phi  < (n+0.5)\Phi _{0}$ with the predominant probability $P _{n} \propto \exp{- E_{n}/ k _{B}T} $, corresponding to the lowest energy $\propto (n - \Phi /\Phi _{0})^{2} $, at almost all measurements \cite{JETP07J,PerMob2001,Letter2003,PCJETP07} corroborate the strong discreteness of the spectrum. Two states $n$ and $n \pm 1$ are observed only in spatial cases \cite{2statesLT34}. But the spectrum discreteness diminishes $ \Delta E_{n,n+1} \approx  I _{p,A} \Phi _{0} = \Phi _{0} q \hbar/mr\overline{(s n _{s})^{-1}}  \rightarrow 0$, Fig.2, at a pair density decrease $ n _{s,A} \rightarrow 0$ in a ring segment $l _{A}$ because $1/\overline{(s n _{s})^{-1}} \approx   s ln _{s,A} n _{s,0}/( l n _{s,A} + l _{A}n _{s,0} - l _{A}n _{s,A}) \rightarrow 0$.

The quantum number $n$ is connected definitely (3) with measurable variables $I _{p}$ and $\Phi $ because of the strong discreteness of the permitted states spectrum (5) which is connected directly with the existence of the phase coherence along the whole ring circuit (2). Therefore it loses a physical sense at $n _{s,A} \rightarrow 0$ when $ \Delta E_{n,n+1} \ll k _{B}T$ and all quantum states have the same probability $P _{n} \propto \exp{- E_{n}/ k _{B}T} $. Thus, the $n$ change, Fig.1, observed at magnetization measurements \cite{nChGeim,nChMoler} should occur through transition between the superconducting states with different connectivity of the wave function, Fig.2. The fundamental difference between these states may be connected with the non-uniqueness of the vector potential $A + \nabla \chi $: $\oint_{l}dl (A + \nabla \chi ) = \oint_{l}dl A = \Phi $ is the gauge-invariant quantity, but $\int_{l}dl A + \int_{l}dl \nabla \chi  \neq \int_{l}dl A $ is not gauge-invariant, where $\chi $ is any continuously differentiable scalar function, for which $\oint_{l}dl \nabla \chi \equiv  0$. The difference between $\oint_{l}dl A $ and $\int_{l}dl A $ and the relation for the canonical momentum $p = mv + eA$ is the essence of the famous Aharonov - Bohm effect \cite{AB1959}. Therefore the quantization phenomena observed in rings are considered \cite{ABRMP1985,ABNanoSt2009,PCScien09} as a case of the Aharonov - Bohm effect. 

The physical sense of the quantum number $n$ and the phase coherence (2) should re-establish when the $l _{A}$ segment will return into the superconducting state $n _{s,A} > 0$. The ring can return to the state with the same or other quantum number depending on the experiment conditions. At the magnetization measurements \cite{nChGeim,nChMoler} superconductivity can break down in a ring segment when the persistent current (4) has reached the critical value $I _{c} = sn _{s}qv _{c} $, i.e. the $|v| $ value has reached the depairing velocity $v _{c} = \hbar /\surd 3\xi (T) m $ \cite{Tinkham}. The pairs velocity equal $v = (\hbar /mr)(n - \Phi /\Phi _{0})$ in a ring with the homogeneous section $s$ and pair density $n _{s}$ along $l$ should reach the depairing velocity $|v| = v _{c} $ at $|n - \Phi /\Phi _{0}| = r/\surd 3\xi (T) > 1/2$ if the ring radius $r$ is larger than the coherence length $\xi (T) $ of the superconductor $r > \surd 3\xi (T)/2 \approx 0.87\xi (T)$. The energy (5) of the $n$ state with $|n - \Phi /\Phi _{0}| > 1/2$ is not lowest. Therefore the quantum number $n$ should change \cite{nChGeim,nChMoler} after the switching between states with different connectivity of the wave function, Fig.2. In the example shown on Fig.1, a segment or whole aluminium ring with the radius $r \approx 0.7 \ \mu m$ and the coherence length $\xi (T) \approx 0.22 \ \mu m $ transits to the normal state at the magnetic flux $\Phi /\Phi _{0} \approx 1.8$; $\Phi /\Phi _{0} \approx 2.8$; $\Phi /\Phi _{0} \approx 3.8$ corresponded to $|n - \Phi /\Phi _{0}| \approx r/\surd 3\xi (T) \approx 1.8$ when $|v| \approx v _{c} $ and returns to the superconducting state after the  jump of  the quantum number from  $n = 0$ to $n = 1$; from $n = 1$ to $n = 2$; from $n = 2$ to $n = 3$ when the pair velocity $v = (\hbar /mr)(n - \Phi /\Phi _{0})$ can have lower value $|v| < v _{c} $ in the state with the closed wave function, Fig.2.  The hysteresis is observed at the magnetization measurements of the rings with $r > 0.87\xi (T)$ and the quantum number $n$ (and consequently $I _{p}$ (4)) may be different at the same $\Phi /\Phi _{0}$ value, depending on the history \cite{nChGeim,nChMoler}. 

Only $n$ states, corresponding to the lowest energy (5), is observed at each $\Phi /\Phi _{0}$ value inside a ring with $r > 0.87\xi (T)$ when the ring or its segment is switched between superconducting and normal states with an external influence or thermal fluctuations. The external influence takes place, for example, at measurements of the critical current $I_{c}$ \cite{JETP07J,PCJETP07,NANO2011} when the external current $I_{ext}$ used for $ I _{c} $ measuring  switches at $I_{ext} = I_{c}$ the ring in the normal state at each measurement, see Fig.2,3 in \cite{PCJETP07}. The thermal fluctuations can switch ring segments in a narrow region $ T _{c} - \delta T _{c}/2 < T < T _{c} +\delta T _{c}/2 $ near superconducting transition $ T _{c} $ where the difference of their energy in normal and superconducting states does not exceed strongly $k _{B}T$ \cite{Tinkham}. $\delta T _{c}$ is the width of the fluctuation region.  Without any external influence the spectrum (5) is discrete, the persistent current (4) $I _{p} \neq 0$ at $\Phi \neq n\Phi _{0} $ and the resistance $R = 0$  permanently below the fluctuation region  $T < T _{c} -\delta T _{c}/2 $. Above this region  $T > T _{c} +\delta T _{c}/2 $ the resistance $R > 0$ and $I _{p} = 0$ also permanently. Therefore under equilibrium condition the persistent current $\overline{I _{p}} \neq 0$ is observed at non-zero resistance $R > 0$ only in the fluctuation region \cite{LP1962,Letter07,toKulik2010,PCScien07}. 

The current circulating in a ring $I(t) = I_{p}\exp -t/\tau_{RL}$ should decay during the relaxation time $\tau_{RL} = L/R$ after the switching of the ring or its segment in the normal state because of a non-zero resistance $R > 0$. The switching can be induced by the external current $I_{ext} > I_{c}$ (or other external influence) at $T < T _{c} -\delta T _{c}/2 $ or thermal fluctuations at $ T _{c} - \delta T _{c}/2 < T < T _{c} +\delta T _{c}/2 $. The experimental results \cite{LP1962,Letter07,toKulik2010,PCScien07,JETP07J,PCJETP07,NANO2011} indicate that the ring reverts with the predominant probability $P _{n} \propto \exp{- E_{n}/ k _{B}T} $ to the state with the same number $n$ at $(n-0.5)\Phi _{0} < \Phi < (n+0.5)\Phi _{0} $ both in the superconducting region at $T < T _{c} -\delta T _{c}/2 $ and in the fluctuation region at $ T _{c} - \delta T _{c}/2 < T < T _{c} +\delta T _{c}/2 $ when all ring segments return to the superconducting state. Therefore both the persistent current and the resistance, measured on average in time, can be together non-zero  $\overline{I _{p}} \neq 0$ at $\overline{R} > 0$. The switching between superconducting states with different connectivity of the wave function, Fig.2, because of the thermal fluctuations explains \cite{PRB2001} the observations of the Little-Parks oscillations of the ring resistance $\Delta R \propto  \overline{I_{p}^{2}}$ \cite{LP1962,Letter07,toKulik2010} and the oscillations of the magnetic susceptibility $\Delta \Phi _{Ip} = L\overline{I_{p}}$ \cite{PCScien07} in the temperature region $ T _{c} - \delta T _{c}/2 < T < T _{c} +\delta T _{c}/2 $ of the resistive transition where $R _{n} > R(T) > 0$ just because of the thermal fluctuations \cite{Tinkham}. Here $R _{n} $ is the ring resistance in the normal state at $T > T _{c} +\delta T _{c}/2 $. The observations of the maximum resistance $\Delta R \propto  \overline{I_{p}^{2}} $ \cite{LP1962,Letter07,toKulik2010} together with $\Delta \Phi _{Ip} = L \overline{I_{p}} = 0$ \cite{PCScien07} at $\Phi = (n+0.5)\Phi _{0} $ testify to the switching of superconducting ring at $T \approx  T _{c}$ between states with different quantum numbers $n$ and $n+1$ because of the impossibility to observe the maximum of $\overline{I_{p}^{2}} = \overline{I _{p,A}^{2}}2\overline{(n - \Phi /\Phi _{0})^{2}}$ and $\overline{I_{p}} = \overline{I _{p,A}}2\overline{(n - \Phi /\Phi _{0})}  = 0$ if only state contributes to these values. The average value of the persistent current is zero $\overline{I_{p}} = 0$ at $\Phi = (n+0.5)\Phi _{0} $ because of the same energy $E_{n+1} = I _{p,A}\Phi _{0} (n +1 - \Phi /\Phi _{0})^{2} = I _{p,A}\Phi _{0}0.5^{2} = E_{n} = I _{p,A}\Phi _{0}(n - \Phi /\Phi _{0})^{2} = I _{p,A}\Phi _{0}(-0.5)^{2}$ (5) and the same probability $P_{n+1} = P_{n}$ of two permitted states $n$ and $n+1$ with opposite $I_{p} $ direction (4): $\overline{I_{p}} \propto  [0.5 + (-0.5) ]/2 = 0$ but $\overline{I_{p}^{2}} \propto  [0.5^{2} + (-0.5)^{2}]/2 = 1/8$. The observations of the periodicity in magnetic field of the resistance $\Delta R \propto  I_{p}^{2}$ \cite{LP1962,Letter07,toKulik2010} and the magnetic susceptibility $\Delta \Phi _{Ip} = LI_{p}$ \cite{PCScien07} are evidence of  the strong discreteness $\Delta E_{n,n+1} \gg k _{B}T$ of the spectrum of superconducting state with closed wave function, Fig.2, even in the region $ T _{c} - \delta T _{c}/2 < T < T _{c} +\delta T _{c}/2 $ where the pair density is not zero $\overline{n _{s}} > 0$ because of thermal fluctuations.

Potential voltage with a direct component $V _{dc}$ can be observed because of this strong discreteness when the same ring segment $l _{A}$ is switched between superconducting and normal states with a frequency $ \omega _{sw} = N _{sw}/\Theta $ by an external influence or thermal fluctuations \cite{JLTP1998}. The potential voltage $ V_{A}(t) = R_{A}I_{p}\exp {-t /\tau_{RL}}$ should appear at each switching in the normal state because of non-zero values  of the $l _{A}$ segment resistance $R_{A} > 0$ and the ring inductance $L$. The direct component, i.e. the voltage $V _{dc} =  \int _{0}^{\Theta }dt V_{A}(t)/\Theta  $ on average in a long time $\Theta  $, equal  $V _{dc} \approx  L\omega _{sw} \overline{I_{p}}$ at $\omega _{sw} \ll  1/\tau_{RL}$ and $V _{dc} \approx R_{A} \overline{I_{p}}$ at $\omega _{sw} \gg  1/\tau_{RL}$, should be non-zero at a non-zero average value of the persistent current  $\overline{I_{p}} \neq 0$. 

The dc voltage oscillations $V _{dc}(\Phi /\Phi _{0}) \propto  \overline{I_{p}}(\Phi /\Phi _{0})$ were observed already on the ring-halves with different sections $s _{w} > s _{n}$ when the asymmetric ring is switched between superconducting and normal states by ac electric current  \cite{Letter2003,PCJETP07} or a noise \cite{Letter07,toKulik2010,PerMob2001}. These observations demonstrate unambiguously  that the persistent current $\overline{I_{p}}$ circulating in the ring clockwise or anticlockwise \cite{PCScien07} can flow against the dc electric field $E = - \bigtriangledown V _{dc}$ directed from left to right or from right to left in the both ring-halves. This paradoxical situation, observed also at measurements of the Little-Parks oscillations \cite{Letter07}, should be connected evidently with the possibility to observe \cite{LP1962,toKulik2010,PCScien07} the persistent current $\overline{I_{p}}$ which does not decay in spite of non-zero resistance $R > 0$ without the Faraday electric field $-dA /dt = 0$. According to the quantum formalism the $\overline{I_{p}} \neq 0$ at $R > 0$ is possible because of the prohibition (4) of the state with  $I_{p} \neq 0$ at $\Phi \neq n\Phi _{0}$ when all ring segments are in superconducting state $n _{s} > 0$ and $1/\overline{(s n _{s})^{-1}} \neq 0$. The quantum formalism demands that the angular momentum of each pair must equal $n\hbar $ when the wave function is closed (2) and, consequently, must change on the value $n\hbar - q\Phi /2\pi = \hbar (n -\Phi /\Phi _{0})$ at the closing of the superconducting state if the pairs velocity $v = 0$ before the closing. This demand as applied to the repeated closing of the wave function with a frequency $ \omega _{sw} = N _{sw}/\Theta $ gives the value 
$$\hbar (\overline{n} -\frac{\Phi }{\Phi _{0}}) \omega _{sw} = rF _{q} \eqno{(6)}$$  
of the angular momentum change in a time unity. The change of the pair momentum in a time unity $F _{q}$ because of the quantization (2) was called in \cite{PRB2001} quantum force. Thus, the inference of the quantum force in \cite{PRB2001} do not overstep the limits of the universally recognised quantum formalism.

\section{Force-free angular momentum transfer}
\label{sec:3}
Hirsch argues \cite{Hirsch2010} {\it "that there is no physical basis for such an azimuthal force"}. I agree with this assertion if only the universally recognised quantum formalism can not be considered as a physical basis. The quantum formalism states, in full accordance with all experimental results, that the persistent current must appear at the closing of superconducting state in the ring, Fig.2, according to (4) and the angular momentum of each pair should change, according to (2), if it was not equal the permitted value $rp = n \hbar $ before the closing. The change $\hbar (n -\Phi /\Phi _{0})$ of the angular momentum of each pair (at $v = 0$ before the closing) must be because of the quantization demand (2). Although the quantum formalism can not explain how this change of macroscopic number $N_ {s} = 2 \pi r sn _{s}$ of pairs can be possible without any force, all experimental results testify to this force-free angular momentum transfer. 

For example, the current $I _{p}$, circulating in the ring, see Fig.1 in \cite{PCJETP07}, when the external current does not exceed the critical current $|I _{ext}| < I _{c+}, I _{c-}$, decays during a relaxation time $\tau_{RL} = L/R $ after the transition into the normal state, see Fig.2 in \cite{PCJETP07}, at $|I _{ext}| = I _{c+}$ or $|I _{ext}| = I _{c-}$. The average velocity of mobile charge carriers falls down to zero $\overline{v} = (I _{p}/sqn_{s})\exp {-t /\tau_{RL}}$ and the angular momentum of electron pair varies from $rp = n \hbar $ to $rp = q \Phi $ because of the dissipation force $F _{dis} = -\eta \overline{v} $ in accordance with the Newton's second law $md\overline{v}/dt = F _{dis}$. Here the relaxation time may be written $\tau_{RL} = m/\eta $. The measurements of the critical current \cite{PCJETP07} corroborate the demand (2) and (4) that the angular momentum and the persistent current must revert to the initial values when the $|I _{ext}|$ decreases down to zero and the ring returns to superconducting state. The oscillations in magnetic field of the critical current $ I _{c+}(\Phi /\Phi _{0})$, $I _{c-}(\Phi /\Phi _{0})$ \cite{JETP07J,PCJETP07,NANO2011} measured through the ring switching between superconducting and normal states are evidence that the persistent current is not zero in the superconducting state. The quantum formalism can describe the oscillations $ I _{c+}(\Phi /\Phi _{0})$, $I _{c-}(\Phi /\Phi _{0})$ measured on the symmetric ring, see Fig.2 in \cite{JETP07J}, and near  $\Phi = n\Phi _{0}$ on the ring with asymmetric link-up of current leads \cite{NANO2011}. The inexplicable discrepancies between experimental results and the theory in the case of asymmetric rings \cite{JETP07J,PCJETP07,NANO2011} puzzle but can not cast doubt on the persistent current in the superconducting state because of the quantum oscillations  $ I _{c+}(\Phi /\Phi _{0})$, $I _{c-}(\Phi /\Phi _{0})$ observed in all cases.

The dissipation force changes the angular momentum of electron pair on $ q\Phi /2\pi - n\hbar  = -\hbar (n -\Phi /\Phi _{0})$ at each transition into the normal state, when, for example, the external current $I _{ext} =  I _{0}\sin (2\pi ft)$, see Fig.1 in \cite{PCJETP07}, with an amplitude $ I _{0} > I _{c+}, I _{c-}$ switch the ring between superconducting and normal states with a frequency $f  \ll 1/\tau_{RL}$. The dissipation force, equal on average in time $\oint _{l}dl \overline{ F _{dis}}/2\pi =  -\hbar (\overline{n} -\Phi /\Phi _{0}) f $, does not change the momentum of mobile charge carriers during a long time $ \Theta \gg 1/f$ because the angular momentum must revert to the permitted value (2) at each return to the superconducting state at $I _{ext} = 0$. The quantum force can provide formally the force balance 
$$\oint _{l}dl \overline{ F _{dis}} + 2\pi rF _{q}=  0 \eqno{(7)}$$
 at the description of this experiment. 

This description should show clearly that the correct statement by Hirsch \cite{Hirsch2010}: {\it "an azimuthal quantum force acting on electrons only would change the total angular momentum of the system, violating the physical principle of angular momentum conservation"} can not concern the quantum force introduced in \cite{PRB2001}. This momentum change in a time unity (6) because of the quantization (2) can provide only formally the force balance (7) at the description of some experiments, like the one considered above, in which the dissipation force is not zero on average in time $\overline{ F _{dis}} \neq 0$. The dissipation force acts between electrons and the crystalline lattice of ions. In order to provide the force balance (7) we should believe that the change of the angular momentum of mobile charge carriers because of the quantization (2) is equilibrated with the same one of ions. Without this balance implied in (7) it could be possible to have rotated a ring merely with help of its switching between superconducting and normal states at $\Phi \neq n\Phi _{0}$ inside it. This possibility can be verified experimentally. I  believe that the ring would not rotate, although this belief implies an additional obvious puzzle. The angular momentum of each superconducting pair should change from $rp = q \Phi /2\pi $ to $rp = n \hbar $ although without a known force but, at least, because of the known cause - its quantization (2). The angular momentum of ions should change even without this cause in order the ring could not rotate. 

\section{Description and explanation}
\label{sec:4}
Hirsch remarks {\it "Nikulov claims that this force explains the Meissner effect as well as the Little-Parks effect"} \cite{Hirsch2010}. I hope it is enough clearly from the above that the quantum force introduced in my paper \cite{PRB2001} does not explain anything but only describes. It is used for description of the Little-Parks effect because thermal fluctuations switch ring segments between superconducting and normal states at $T \approx  T _{c}$, but it is useless for a description of the Meissner effect. The azimuthal quantum force could be introduced for the description of an experiment when a bulk superconductor placed in a weak magnetic field is switched repeatedly with a frequency $ \omega _{sw} = N _{sw}/\Theta $ between superconducting and normal states. The dissipation force will change the angular momentum of mobile charge carriers after each transition into the normal state as well as in the case of the ring considered above. The quantum force $rF _{q} = \hbar (-\Phi /\Phi _{0}) \omega _{sw}$, see (6), may be introduced only in this case in order to provide nominally the force balance (7) on average in time. The angular momentum of each pair can change at the Meissner effect from $rp = q \Phi /2\pi $ to $rp = 0 $ on a macroscopic value $ \hbar (- \Phi /\Phi_{0}) = \hbar (-B \pi r^{2}/\Phi_{0})  $ in contrast to the case considered above when this change can not exceed the Planck's constant $ \hbar (n - \Phi /\Phi_{0}) \leq  \hbar \times 0.5$ irrespective of the ring radius $r$ and the magnetic flux $\Phi = B\pi r^{2}$ value. For example, at the expulsion of a magnetic field $B \approx  0.01 \ T$ from a superconducting cylinder with a radius $r \approx  0.01 \ m$ and a height $h \approx  0.1 \ m$ the angular momentum of each pair should change on $ \hbar (- \Phi /\Phi_{0}) \approx  \hbar \times  10^{9}$ and  all $N_{s} = n_{s}\pi r^{2}h \approx  10^{23}$ pairs (at a typical density $ n_{s} \approx  10^{28} \ m^{-3}$) on $ \hbar \times 10^{31} $. It is in truth macroscopic puzzle which the azimuthal quantum force \cite{PRB2001} can not explain and even describe.  

This quantum force (6) can be useful for a description only of phenomena or experiments with recurring switching between superconducting and normal states with a frequency $ \omega _{sw} = N _{sw}/\Theta $. In the case of the Little-Parks and other such effects \cite{LP1962,Letter07,toKulik2010,PCScien07} the quantum force replaces the force $-qdA/dt$ of the Faraday electric field $-dA/dt$ in order to describe the persistent current $\overline{I _{p}} \neq 0$ observed at non-zero resistance $\overline{R} > 0$ on average in time. This azimuthal force $2\pi rF _{q}$ maintains 
$$\frac{2\pi rF _{q}}{q}  = \overline{RI} \eqno{(8)}$$ 
the current $\overline{I _{p}}$ circulating in a ring with a resistance $\overline{R} > 0$, as well as the Faraday's electromotive force makes this in accordance with the Ohm's law $-2\pi rdA/dt = RI$. The relation (8) could be interpreted as an analogue of the Ohm's law if only the relation $\overline{RI} = \overline{R} \times \overline{I} $ can be correct. The measurements of the quantum oscillations of the resistance \cite{LP1962,Letter07,toKulik2010} and of magnetic susceptibility at $\overline{R}$ \cite{PCScien07} are evidence that both $\overline{I _{p}} \neq 0$ and $\overline{R} > 0$ in the fluctuation region $ T _{c} - \delta T _{c}/2 < T < T _{c} +\delta T _{c}/2 $. It is possible because the ring or its segments can be in the normal state with $R _{n} > 0$ during an average time $t _{n}$ and in superconducting state with $I _{p} \neq 0$ during other time $t _{s}$ between switching. The persistent current and the resistance measured in \cite{Letter07,toKulik2010} can be estimated with the relations  $\overline{R} \approx  R _{n} t _{n}\omega _{sw} $, $\overline{I _{p}} \approx I _{p} t _{s}\omega _{sw} $, whereas $\overline{RI} \approx  \omega _{sw} L I _{p}$ at $\omega _{sw} \ll  \tau_{RL}$ because the current and the resistance can be non-zero at the same time only during the relaxation time $\tau_{RL} $. The relation (8) describes only the transitional processes between the current $I _{p}$ determined by the quantization (4) and by the dissipation and therefore $\overline{RI} \neq \overline{R} \times \overline{I} $ in the common case. The approximate equality $\overline{RI} \approx  \overline{R} \times \overline{I} $  is possible only in the limit case  $\omega _{sw} \gg  \tau_{RL}$. Nevertheless the observations \cite{LP1962,Letter07,toKulik2010,PCScien07} $\overline{I _{p}} \neq 0$ at $\overline{R} > 0$   can not be described without (8) because the transitional processes must be between the states  $I _{p} \neq 0$, $R = 0$ and $|I| < |I _{p}| $, $R > 0$.

The paradoxical observations \cite{Letter07,toKulik2010} of the current $\overline{I _{p}} $  flowing against the force of electric field $E = - \bigtriangledown V $ can not be described also without the quantum force. The Little-Parks  oscillations \cite{LP1962} are observed as a rule \cite{Letter07,toKulik2010} with help of measurement of the dc voltage $V = RI _{ext}$ induced by an external current flowing from left to right (or from right to left) through the ring-halves, see Fig.1 in \cite{Letter07}. The voltage periodicity $V(\Phi /\Phi_{0}) = R(\Phi /\Phi_{0})I _{ext}$, see, for example, Fig.3 in \cite{Letter07}, is evidence of the persistent current $\overline{I _{p}} \neq 0$ at $\Phi \neq n\Phi_{0}$ flowing against the electric field  $\overline{E} = - \bigtriangledown \overline{V} $ on average in time in one of the ring-halves. The voltage $\overline{R} I _{ext} \approx  R _{n} t _{n}\omega _{sw} I _{ext}$ can be observed because of non-zero resistance during a time $t _{n}$. Mobile charge carriers can move against the electric field only during the relaxation time under its own inertia (kinetic inductance) in accordance with the Newton's second law $md\overline{v}/dt = F _{dis} - qE$. But it is possible only at their non-zero velocity corresponding to the permitted state (2) before the transition into the normal state, i.e. thanks to angular momentum change because of the quantization. The dc voltage $V _{dc}(\Phi /\Phi _{0}) \propto  \overline{I_{p}}(\Phi /\Phi _{0})$, which should be induced with segment switching $V _{dc} \approx  L\omega _{sw} \overline{I_{p}}$ and were observed \cite{Letter07,toKulik2010}, can not be described also without the quantum force. It can equilibrate 
$$\frac{ l _{A}F _{q}}{q}  = \overline{RI} - V _{dc}; \ \ \ \frac{ (l - l _{A})F _{q}}{q}  = V _{dc} \eqno{(9)}$$
the dc voltage $V _{dc}$ which should be observed both on the switched segment $l _{A}$ and the superconducting segment  $l - l _{A}$

I follow blindly in this paper and in \cite{PRB2001} to the orthodox quantum mechanics which rather describes phenomena than explains why these phenomena can be observed. The Meissner effect and the persistent current $\overline{I _{p}} \neq 0$ observed at $R > 0$ are not only phenomena which were not explained. The most well-known example is the double-slit interference experiment \cite{electron,neutron,atom}. Other one is the Stern-Gerlach effect \cite{SternGerlach}. Quantum mechanics describes these phenomena but it refuses to explain how a particle can make its way through two slits at the same time and how the magnetic moment of electron or atom can have the same value of projection on any direction. Some other issues, which quantum mechanics refuses to address, are considered in the book \cite{QuCh2006}. This refusal to search  causes of quantum phenomena is based upon epistemological belief, known as instrumentalism or positivism, which explicitly denies any explanatory role of science. According to this point of view a physical theory should only provide a computational instrument for description and prediction of experimental results. Concerning this epistemological problem the founding fathers of quantum theory can be divided into positivists, Heisenberg, Bohr, Born, Dirac, Pauli and realists, Planck, Einstein, Schrodinger, de Broglie. The realists, for example Einstein, stated that quantum mechanics can not be interpreted as the complete theory because of its repudiation of {\it "the programmatic aim of all physics: the complete description of any (individual) real situation (as it supposedly exists irrespective of any act of observation or substantiation)"} \cite{Einstein1949}. The positivists, on the other hand, for example Heisenberg, stated that {\it "Any attempt to find such a description would lead to contradictions"} and therefore {\it "the term 'happens' is restricted to the observation"} \cite{Heisenberg1958}.

I would like to be among realists. But my azimuthal quantum force \cite{PRB2001} does not overstep the limits of the positivism of the orthodox quantum mechanics whereas Hirsch \cite{Hirsch2010} attempts to explain what happens at the Meissner effect. His explanation may seem strange. But the famous interpretation of the quantum theory in terms of hidden variables by David Bohm \cite{Bohm1952} and hidden variables model by John Bell \cite{Bell1966} seem even more strange. The paradoxicality of some quantum phenomena rules out the possibility of no paradoxical explanation. The magnetic Lorentz force acting on the radially outgoing charge, assumed by Hirsch \cite{Hirsch2010}, seems to be only way to explain the Meissner effect without challenge to the principle of angular momentum conservation. But this way can not be valid for an explanation of the angular momentum change $ \hbar (n -\Phi /\Phi _{0})$ at the closing of superconducting state in the ring, Fig.2, because of the obvious impossibility of an radial current. Quantum mechanics predicts but can not explain also a real mechanical force which can act and be measured at mechanical closing of the superconducting loop at $\Phi \neq n\Phi _{0}$, see the end \cite{PRB2001}. An unprejudiced consideration of these and other puzzles generated by the orthodox quantum mechanics would clarify its essence. 

\section{What do the puzzles reveal about the orthodox quantum mechanics?}
\label{sec:5}
The outright refusal to raise some questions is debated from the very outset of quantum mechanics and up to now. The main point at issue may be connected with two puzzles of quantum phenomena: wave-particle duality and indeterminism. Einstein, who introduced into the consideration both the wave-particle duality in 1905 \cite{Einstein1905} and indeterminism in 1916 \cite{Einstein1916}, understood in full measure the paradoxicality of these features. Bohr remembers in \cite{Bohr1949} picturesque phrase by Einstein about {\it "ghost waves (Gespensterfelder) guiding the photons"} and Einstein himself wrote in 1954: {\it "All these fifty years of conscious brooding have brought me no nearer to the
answer to the question, 'What are light quanta?' Nowadays every Tom, Dick and Harry thinks he knows it, but he is mistaken."} Einstein, as well as some other experts later on \cite{Lamb95}, could not accept the intrinsic inconsistency of the concept of light quanta, which he proposed: the photon with a energy $E = h\nu $ and momentum $p = h/\lambda $ is, on the one hand, the wave with the certain frequency $\nu $ and length $\lambda $ which can not be localised, but on the other hand it must interact with a localised object, for example an atom. Bohr wrote: {\it "The acuteness of the dilemma is stressed by the fact that the interference effects offer our only means of defining the concepts of frequency and wavelength entering into the very expressions for the energy and momentum of the photon"} \cite{Bohr1949}. The orthodox quantum mechanics describes this dilemma with help of the principle of states superposition or, equivalently, of the positivistic interpretation of the Schrodinger wave function proposed by Born. The superposition of states, considered \cite{LandauL} as the cardinal positive principle of quantum mechanics, must imply its collapse at observation because of the logical impossibility to see anything in some states simultaneously.

The concept of the collapse, introduced by von Neumann as far back as 1932 \cite{Neumann1932}, {\it remains most puzzling and counterintuitive aspect of the interpretation of quantum mechanics} \cite{Cramer1986}. The well-known puzzles generated with this concept are revealed with two famous paradoxes, of "Schrodinger's cat" \cite{Schrod35D} and the EPR paradox \cite{EPR1935}. The first puzzle indicates a fuzzy status of measurements in quantum mechanics criticised by John Bell \cite{Bell1987,Bell1990} and discussed up to now \cite{Mermin2006,Ghirardi2008,Mermin2008}. The EPR paradox reveals, by the use of the most paradoxical quantum principle, the "entanglement" \cite{Schrod35D,Schrod35E} or the "EPR correlation" \cite{EPR1935}, the contradiction of the concept of superposition and its collapse with local realism. This contradiction is one of the most discussed problems in the last years \cite{Norsen09,Ghirardi10} thanks to the famous Bell's theorem \cite{Bell1964} which has showed that the EPR paradox led to experimentally testable differences between quantum mechanics and local realistic theories. Bell, as well as some other experts \cite{Norsen09,Ghirardi10}, interpret experimental evidence \cite{Aspect1981} of violation of the famous Bell's inequality as {\it "the real problem with quantum theory: the apparently essential conflict between any sharp formulation and fundamental relativity"}. Because of this essential conflict revealed first by Einstein as far back as 1927 \cite{Einstein1927} the adherents of the orthodox quantum mechanics are forced to state that the wave function can describe only knowledge of the quantum system but not the system in itself, see \cite{Cramer1986,Mermin2001}, for example.

The knowledge-information interpretation is enough natural for the positivistic interpretation of the Schrodinger's wave function $\Psi (r)$ proposed by Born. According to this interpretation $|\Psi (r)|^{2}dV$ is a probability to observe a particle in an element of volume $dV$. Before the observation the probability $|\Psi (r)|^{2}dV$ can have a finite value $|\Psi (r)|^{2}dV < 1$ for all elements $dV$ of a large volume because we can not know where the particle will be found. But after the observation  the probability is $|\Psi (r)|^{2}dV = 1$ in the one element, where the particle was observed, and  $|\Psi (r)|^{2}dV = 0$ for all others. The wave function $\Psi (r)$, describing superposition of states before the observation, collapses after the observation because of the change of our knowledge. {\it The collapse must take place instantaneously over all space} \cite{QuCh2006}, or rather it must be outside of the real physical time. This non-realistic interpretation seems to have no alternative for the description of some quantum phenomena, for example, the double-slit interference experiments \cite{electron,neutron,atom}. But it is obviously not valid for the description of superconductivity phenomena. 

The GL wave function $\Psi _{GL} = |\Psi _{GL}|\exp i\varphi $ describes quite real density of superconducting pairs in accordance with {\it the programmatic aim of all physics} \cite{Einstein1949} upheld by Einstein and other realists. The real density $|\Psi _{GL}|^{2} = n _{s} $ can not collapse at the act of observation and can not depend on our knowledge. The GL wave function $\Psi _{GL} = |\Psi _{GL}|\exp i\varphi $ can alter only because of a real physical influence, for example, a heating of the $l _{A}$ ring segment, Fig.2. Because of this fundamental difference between the positivistic Born's interpretation of the Schrodinger's wave function and the realistic essence of the GL wave function, the puzzles revealed by Hirsch \cite{Hirsch2010} and the quantum force \cite{PRB2001} differ basically from the puzzles revealed by Schrodinger with his cat paradox \cite{Schrod35D}, by Einstein with the EPR paradox \cite{EPR1935}, by Bell \cite{Bell1966,Bell1964} with his no-hidden-variables theorems \cite{Mermin1993} and by other adherents of realism \cite{QuCh2006}. The realists can have no reason for complaint of the orthodox description of superconductivity phenomena because of its realistic essence. The sharp criticism by Bell \cite{Bell1987,Bell1990} and others \cite{QuCh2006} of the fuzzy status of measurements in quantum mechanics bears no relation to this description. There is not a measurement problem at the description of macroscopic quantum phenomena. One of the puzzles of the measurement problem raised by Bell is this: {\it "how exactly is the world to be divided into speakable apparatus...that we can talk about...and unspeakable quantum system that we can not talk about?"} \cite{Bell1984}. This puzzle is generated with the necessity to amplify {\it microscopic events to macroscopic consequences} \cite{Bell1984}. This necessity is absent for macroscopic events. Therefore macroscopic quantum systems are {\it speakable} in contrast to the microscopic one. 

Bell urged to study {\it the de Broglie-Bohm picture} which {\it "disposes of the necessity to divide the world somehow into system and apparatus"} \cite{Bell1984}. He marked out as far back as 1964 \cite{Bell1964} the theory by David  Bohm \cite{Bohm1952} as the evidence of the possibility to interpret the quantum theory realistically in terms of hidden variables. This {\it Bell's favorite example of a hidden-variables theory is not only explicitly contextual but explicitly and spectacularly non-local, as it must be in view of the Bell-KS theorem and Bell's Theorem"} \cite{Mermin1993}. The non-locality of the quantum mechanics is revealed by Bohm \cite{Bohm1952} by means of a "quantum mechanical" potential. The Bohm's quantum potential is deduced \cite{Bohm1952} in the limits of the universally recognised quantum formalism, as well as the azimuthal quantum force in superconductor \cite{PRB2001}. But there is a fundamental difference between the Bohm's quantum potential and the quantum force \cite{FPP2008} because of the different essence of the wave functions \cite{QTRF2007} describing different quantum phenomena. 

This difference may be seen clearly on the example \cite{FPP2008} of the Aharonov - Bohm effect \cite{AB1959} observed in the double-slit interference experiment and superconducting ring. The effect in the both cases is a consequence of the connection $\hbar \bigtriangledown \varphi = p = mv + qA$ between the phase $\varphi $ of the wave function $\Psi = |\Psi |\exp i\varphi $ and the vector potential $A$. But the essence of $|\Psi |^{2}$ is positivistic at the description of the double-slit interference experiment and realistic at the description of the Aharonov - Bohm effect in the ring \cite{FPP2008}. The quantization (2) is a consequence just of the realistic essence of the wave function without which the requirement $\Psi  = |\Psi |\exp i\varphi = |\Psi |\exp i(\varphi + n2\pi)$ can not be valid. The phase difference $\varphi _{1} - \varphi _{2}$ for two possible path $l _{1}$, $l _{2}$ of a particle through the first slit $\varphi _{1} = \int_{S}^{y} dl _{1} \nabla \varphi $ and the second slit $\varphi _{2} = \int_{S}^{y} dl _{2} \nabla \varphi $ between a particle source $S$ and a point $y$ on the detecting screen, see Fig.1 in \cite{FPP2008}, equals the integral $ \oint_{l}dl \nabla \varphi $ in (2). But the value $\varphi _{1} - \varphi _{2} = \oint_{l}dl \nabla \varphi = \Delta \varphi _{0} + 2\pi \Phi /\Phi_{0} $ changes uninterruptedly, in defiance of (2), with the co-ordinate $y$ and magnetic flux $\Phi $ because of the collapse of the wave function $\Psi  = |\Psi _{1}|\exp i\varphi (l _{1})+ |\Psi _{2}|\exp i\varphi (l _{2})$, describing the double-slit interference experiment, at observation of a particle position $y$ on the detecting screen. This positivistic cause of the quantization (2) breach differs in essence from the real rupture of superconducting state $|\Psi _{GL}|^{2}$ in the ring, Fig.2, under influence of a real cause. The quantum force (6) may be considered real just because of the reality of  the $|\Psi _{GL}|^{2}$ change.

\section*{Conclusion}
\label{sec:6}
Although the azimuthal quantum force is deduced in the limits of the universally recognised quantum formalism as well as the Bohm's quantum potential \cite{Bohm1952} it is not useless. Heisenberg stated \cite{Heisenberg1958} that according to the pure positivistic point of view the Bohm's interpretation \cite{Bohm1952} repeats only the orthodox Copenhagen interpretation. Bell did not contest this statement but the Bohm's work \cite{Bohm1952} inspired he, as the opponent of positivism, with the confidence in the possibility of realistic interpretation of quantum mechanics. Bell was sure that variables in any realistic theory should be {\it "'hidden' because if states with prescribed values of these variables could actually be prepared, quantum mechanics would be observably inadequate"} \cite{Bell1966}. But all superconducting states with prescribed values of these variables can actually be prepared. These variables, including the angular momentum change $ \hbar (n -\Phi /\Phi _{0})$ at the closing of superconducting state in the ring, Fig.2, are real in the sense that they can not change because of an act of observation as such. The realistic description of superconductivity phenomena even without hidden variables removes the puzzles generated with superposition collapse at observation but other puzzles become more real in this description. The problem of the force-free momentum transfer at the Aharonov - Bohm effect \cite{QuCh2006}, discussed many years \cite{AB1961,Book1989,Boyer00,PRL07,Nature08} for the case of the double-slit interference experiment, is more real at the closing of the superconducting state in the ring, Fig.2, because the wave function $\Psi _{GL} = |\Psi _{GL}|\exp i\varphi $ describes the real density $|\Psi _{GL}|^{2} = n_{s}$. The EPR paradox and the Bohm's quantum potential \cite{Bohm1952} have revealed the non-locality of the quantum mechanics in description (non-locality of the first kind according to \cite{Cramer1986}) and the Bell's no-hidden-variables theorem \cite{Bell1964} has revealed this non-locality in observation (non-locality of the second kind \cite{Cramer1986}). This non-local correlation because of the instantaneous collapse of the positivistic wave function at observation is absent at the description of superconductivity phenomena because of its realistic essence. Nevertheless there is a non-locality, at least in description. The orthodox quantum formalism predicts that the velocity of pairs in a superconducting segment $l_{B}$ of the ring, Fig.2, should change without a real force at the transition of the spatially separated segment $l_{A}$ in superconducting state. This change must be because of the quantization requirement (2) but the question how quickly the state of pairs can change in the spatially separated segments was essentially never raised nor answered. This change must not be instantaneous, as in the case of the collapse, but the quantum mechanics or any other theory can not provide a defined time or velocity of this real influence on the spatially separated event.

The benefit of the quantum force considered in \cite{PRB2001} and this paper is not limited to the disclosure of new puzzles generated with quantum mechanics. First of all, the mysterious change of the angular momentum at the closing of the wave function (6) delivers from the necessity to make preposterous claim that an electric current can be dissipationless in a ring with non-zero resistance. This claim, made, for example by the authors \cite{PCScien09,Birge2009}, is not only fully groundless but also useless \cite{toKulik2010} because of the observation \cite{Letter07} of the persistent current flowing against electric field. The authors \cite{PCScien09} admit that {\it "A dissipationless equilibrium current flowing through a resistive circuit is counterintuitive"} but claim unreasonably that {\it "it has a familiar analog in atomic physics: Some atomic species' electronic ground states possess nonzero orbital angular momentum, which is equivalent to a current circulating around the atom"}. The falsity of this analogy is obvious from the experimental results \cite{PCScien09} of these authors. They observe the one-dimensional angular momentum ${\bf M _{p}} = 0 \times  {\bf i _{x}} + 0 \times  {\bf i _{y}} + (2m/e) I _{p} S \times  {\bf i _{z}}$  of the persistent current in flat rings the direction of which changes periodically with magnetic flux value. (It is obvious that the projections of this angular momentum in the ring flat $x \times  y$ are absent,  $M _{p,x} = 0$, $M _{p,y} = 0$.) The angular momentum of atom should be considered as three-dimensional  ${\bf s} = s_{x} \times  {\bf i _{x}} + s_{y} \times  {\bf i _{y}} + s_{z} \times  {\bf i _{z}}$ if its direction could be observed, i.e. if three its projections $s_{x}$, $s_{y}$, $s_{z}$ could be measure simultaneously. Because of the impossibility of such measurement the angular momentum of atom can not exist really according to the orthodox quantum mechanics and can be considered only as hidden variable in realistic theories \cite{Bell1966}. 

The exponential decrease of the amplitude of the persistent current with temperature increase observed in normal metal rings, see Fig.3 in \cite{PCScien09}, makes absolutely nonsensical both the analogy with atom and the claim \cite{PCScien09,Birge2009} on the absence of dissipation. The experimental dependencies, Fig.3 in \cite{PCScien09}, are described enough well with the theory considered diffusive electrons the mean free path of which is small because of scattering on different kinds of disorders of realistic metal rings. Such diffusive motion of electron and its scattering on disorders are absolutely inconceivable on a stationary atomic orbit. The claim \cite{PCScien09,Birge2009} that the persistent current can flow at $R > 0$ without dissipating energy implies the absence of its interaction with environment and, consequently and contrary to the experiment, Fig.3 in \cite{PCScien09}, the impossibility of any temperature dependence of its parameters. The mysterious observations \cite{PCScien09} of the persistent current which does not decay in the realistic metal rings can be described without this preposterous claim if one takes into account that electron can return after its scattering in the state with certain angular momentum described by the quantization (2). This returning from the state with uncertain momentum and zero velocity on the average to the quantization state with certain momentum and non-zero velocity on the average at $I _{p} \neq 0$ should imply the momentum change opposite to the one because of the electron scattering. The second change is described by the dissipation force and the first one may be called as quantum force.   

This description with the quantum force withstanding the dissipation force may predict a possibility of the potential difference $V_{dc}(\Phi /\Phi_{0}) \propto I_{p}(\Phi /\Phi_{0})$ on halves of asymmetric normal metal ring with the persistent current. The observations of this phenomenon on halves of asymmetric superconducting rings \cite{Letter07,toKulik2010} testify to the validity of this prediction. In order that this prediction could be verified by an experimenter he should reject the preposterous claims \cite{PCScien09,Birge2009} about the dissipationless current at $R > 0$ and the analogy with atom. This  analogy does not take into account not only the obvious fundamental difference between atom and mesoscopic ring but also the difference of our experimental possibilities on these different levels. We can not make an asymmetric atom, but it is enough easy to make asymmetric ring. This difference of the experimental possibilities is more important than the non-universality in the quantum description because physics is empirical science and quantum mechanics describes first of all phenomena. One can switch with a frequency $\omega _{sw}$ the segment $A$, Fig.2, between superconducting and normal state with help, for example, of laser beam and measure the voltage induced with this switching. One can also measure the mechanical force acting at mechanical closing of the superconducting loop at $\Phi \neq n\Phi _{0}$ \cite{PRB2001}. The orthodox quantum mechanics can describe even these phenomena but it can not explain why the angular momentum can change because of the closing of the wave function. We can only hope or dream that this puzzle will be solved sometime.

\end{document}